\documentstyle[12pt]{article}
\newcommand{\lsim}{\raisebox{0.3mm}{\em $\, <$}
\hspace{-3.3mm} \raisebox{-1.8mm}{\em $\sim \,$}}
\newcommand{\gsim}{\raisebox{0.3mm}{\em $\, >$}
\hspace{-3.3mm} \raisebox{-1.8mm}{\em $\sim \,$}}
\begin{document}
\rightline{hep-ph/9606411}
\rightline{TMUP-HEL-9605}
\rightline{June 1996}
\baselineskip=19pt
\vskip 0.7in
\begin{center}
{\large{\bf A STERILE NEUTRINO SCENARIO}}
{\large{\bf CONSTRAINED BY EXPERIMENTS AND COSMOLOGY}}
\end{center}
\vskip 0.4in
\begin{center}
Nobuchika Okada
\footnote{JSPS Research Fellow}
\footnote{Email: n-okada@phys.metro-u.ac.jp}
and Osamu Yasuda\footnote{Email: yasuda@phys.metro-u.ac.jp}

\vskip 0.2in
{\it Department of Physics, Tokyo Metropolitan University}

{\it 1-1 Minami-Osawa Hachioji, Tokyo 192-03, Japan}
\end{center}

\vskip .7in
\centerline{ {\bf Abstract} }

We discuss a model in which three active and one sterile neutrino
account for the solar, the atmospheric and the LSND neutrino
anomalies.  It is shown that if $N_\nu<4$ then these and other
experiments and big bang nucleosynthesis constrain all the mixing
angles severely, and allow only the small-angle MSW solution.  If
these neutrinos are of Majorana type, then negative results of
neutrinoless double beta decay experiments imply that the total mass
of neutrinos is not sufficient to account for all the hot dark matter
components.
\newpage

\section{Introduction}
Recently LSND group \cite{lsnd} (See also \cite{hill})
reported that they have found candidate events for
$\overline \nu_\mu\rightarrow\overline \nu_e$ oscillation.
Combing their result in \cite{lsnd} with the data by
the E776 group \cite{e776} on the same channel
$ \nu_\mu\rightarrow \nu_e$ and the reactor data by Bugey \cite{bugey},
the mass squared difference $\Delta m^2$
seems to be in the range
\begin{eqnarray}
0.27{\rm eV}^2\lsim
\Delta m_{{\rm LSND}}^2
\lsim 2.3{\rm eV}^2.
\label{eqn:delmlsnd}
\end{eqnarray}
On the other hand, it has been argued
that the solar neutrino problem
\cite{homestake}--\cite{sage}
is solved if a set of the oscillation
parameters of the two-flavor neutrino mixing satisfies one of the followings
\cite{solar1}--\cite{solar5}:
\begin{eqnarray}
&{ }&(\Delta m^2_\odot,\sin^22\theta_\odot)\nonumber\\
&\simeq&\left\{ \begin{array}{lr}
({\cal O}(10^{-10}{\rm eV}^2),{\cal O}(1)),&
({\rm vacuum~oscillation~solution})\\
({\cal O}(10^{-5}{\rm eV}^2),{\cal O}(10^{-2})),&
({\rm small-angle~MSW~solution})\\
({\cal O}(10^{-5}{\rm eV}^2),{\cal O}(1))&
({\rm large-angle~MSW~solution}).
	     \end{array} \right.
\label{eqn:solar}
\end{eqnarray}
Furthermore, it has been reported by the Kamiokande group
\cite{kamioka1}\cite{kamioka2} that their
atmospheric neutrino data suggests neutrino oscillation
with a set of parameters $(\Delta m^2,~\sin^22\theta)\simeq(1.8\times10^{-2}
{\rm eV}^2,1.0)$ for $(\nu_e\leftrightarrow\nu_\mu)$, ($1.6\times10^{-2}
{\rm eV}^2,1.0)$ for $(\nu_\mu\leftrightarrow\nu_\tau)$
(See also \cite{imb}--\cite{soudan2}).

The results of experiments on these anomalies have been given in the
framework of two-flavor mixings in the original literatures, and have
been analyzed from the viewpoint of the three flavor mixing by many
people \cite{kp}-\cite{flm}.  It has been shown that
strong constraints are obtained if two of the three anomalies are
taken for granted, and it seems difficult \cite{ym} (See also
\cite{cf1}) to account for all these three anomalies within the three
flavor mixing.

Here we investigate the possibility in which one sterile neutrino as
well as three known flavors of neutrinos are responsible for these
anomalies.  Some features of this possibility have been discussed in
the past
\cite{pv}--\cite{p}.
In this paper we analyze in detail the mass squared differences and
the mixing angles in this scenario.  It turns out that if the number
$N_\nu$ of light neutrinos is less than 4, then because of strong
constraints by the solar and atmospheric neutrino observations,
accelerator and reactor experiments including LSND as well as big bang
nucleosynthesis, all the 6 mixing angles are strongly constrained.  In
this case 4$\times$4 mixing matrix is effectively split into
2$\times$2 $\nu_e\leftrightarrow\nu_s$ and
$\nu_\mu\leftrightarrow\nu_\tau$ matrices, and only the small-angle
MSW solution is allowed.  If we assume that these neutrinos are of
Majorana type, then the upper bound on $\langle m_{\nu_e}\rangle$ from
neutrinoless double $\beta$ decay experiments suggests that these
neutrinos are not heavy enough to explain all the hot dark matter
components.  If $N_\nu\ge 4$, then all the solutions to the solar
neutrino problem are allowed, hot dark matter can be accounted for by
neutrinos and there may be a chance to observe neutrinoless double
$\beta$ decays in the future experiments.

It has been pointed out
that physics of supernova gives constraints on sterile neutrinos
\cite{qfmmw}\cite{kmp}, but these constraints apply only to $\Delta
m^2$ which is larger than the mass scale suggested by the LSND data,
so we will not discuss this point in this paper.

In section 2 we present our formalism on oscillations among four
species of neutrinos.  In section 3 we discuss constraints from
reactor and accelerator experiments.  In section 4 constraints by big
bang nucleosynthesis are considered.  In section 5 we give constraints
from neutrinoless double $\beta$ decay experiments.  In section 6 we
examine the possibility that neutrinos are hot dark matter.  In
section 7 we discuss consequences in case of $N_\nu\ge 4$.
In section 8 we give our conclusions.

\section{Oscillations Among Four Flavors}

Let us consider a model with three flavors of neutrinos
$\nu_e,~\nu_\mu,~\nu_\tau$ and one sterile neutrino $\nu_s$,
which is singlet with respect to all the gauge groups in the standard
model.  We assume that ($\nu_e,~\nu_\mu,~\nu_\tau,~\nu_s$) are related to
the mass eigenstates ($\nu_1,~\nu_2,~\nu_3,~\nu_4$) by the following
unitary matrix:
\begin{eqnarray}
U&\equiv&\left(
\begin{array}{cccc}
U_{e1} & U_{e2} &  U_{e3} &  U_{e4}\\
U_{\mu 1} & U_{\mu 2} & U_{\mu 3} & U_{\mu 4}\\
U_{\tau 1} & U_{\tau 2} & U_{\tau 3} & U_{\tau 4}\\
U_{s1} & U_{s2} &  U_{s3} &  U_{s4}
\end{array}\right)\nonumber\\
&\equiv&e^{i\alpha'}e^{i\beta'\lambda_3}e^{\sqrt{3}i\gamma'\lambda_8}
e^{\sqrt{6}i\delta'\lambda_{15}}V_{KM}
e^{-\sqrt{6}i\gamma\lambda_{15}}e^{-\sqrt{3}i\beta\lambda_8}
e^{-i\alpha\lambda_3}.
\label{eqn:u}
\end{eqnarray}
Here
\begin{eqnarray}
V_{KM}&\equiv& R_{34}({\pi \over 2}-\theta_{34})R_{24}(\theta_{24})
e^{i\delta_1\lambda_3}R_{23}({\pi \over 2}-\theta_{23})
e^{-i\delta_1\lambda_3}\nonumber\\
&{ }&e^{i\delta_3\lambda_{15}}
R_{14}(\theta_{14})e^{-i\delta_3\lambda_{15}}
e^{i\delta_2\lambda_8}R_{13}(\theta_{13})e^{-i\delta_2\lambda_8}
R_{12}(\theta_{12})
\label{eqn:vkm}
\end{eqnarray}
is a $4\times4$ Kobayashi-Maskawa matrix for the lepton sector,
$c_{ij}\equiv\cos\theta_{ij}$,$s_{ij}\equiv\sin\theta_{ij}$,
\begin{eqnarray}
R_{jk}(\theta)\equiv \exp\left(iT_{jk}\theta\right)
\end{eqnarray}
is a $4\times4$ orthogonal matrix with
\begin{eqnarray}
\left(T_{jk}\right)_{\ell m}=i\left(\delta_{j\ell}\delta_{km}
-\delta_{jm}\delta_{k\ell}\right),
\end{eqnarray}
$2\lambda_3\equiv{\rm diag}(1,-1,0,0)$,
$2\sqrt{3}\lambda_8\equiv{\rm diag}(1,1,2,0)$,
$2\sqrt{6}\lambda_{15}\equiv{\rm diag}(1,1,1,-3)$
are diagonal elements of the $su(4)$ generators.

The four phases $\alpha',~\beta',~\gamma',~\delta'$ in front of
$V_{KM}$ in (\ref{eqn:u}) can be absorbed by redefining the wave
functions of charged leptons.  If there is a Majorana mass term,
however, the three phases $\alpha,~\beta,~\gamma$ cannot be absorbed
\cite{bhp}--\cite{fy}.  Since these factors $\alpha,~\beta,~\gamma$
are cancelled in the probability of neutrino oscillations, they do not
affect the results of neutrino oscillations, but as we will see later,
they do affect the effective mass of $\nu_e$ and the constraint on the
masses of neutrinos from neutrinoless double $\beta$ decay.

We can assume without loss of generality that
\begin{eqnarray}
m_1^2<m_2^2<m_3^2<m_4^2.
\label{eqn:inequality}
\end{eqnarray}
Three mass scales $\Delta m_\odot^2\sim{\cal O}(10^{-5}{\rm eV}^2)$ or
${\cal O}(10^{-10}{\rm eV}^2)$, $\Delta m_{\rm atm}^2\sim{\cal O}
(10^{-2}{\rm eV}^2)$, $\Delta m_{\rm LSND}^2\sim{\cal O}(1{\rm eV}^2)$
seem to be necessary to explain the suppression of the ${}^7$Be solar
neutrinos \cite{solar1}, the zenith angle dependence of the Kamiokande
multi-GeV data of atmospheric neutrinos \cite{kamioka2}, and the LSND
data \cite{lsnd},\footnote{It can be shown that the probability of
neutrino oscillation in vacuum in the framework of more than three
flavors is reduced essentially to the two flavor case, once one
assumes the mass hierarchy \cite{3nu}\cite{bgk}, so that the analysis
of \cite{lsnd} indicates that the mass squared difference in our case
should satisfy (\ref{eqn:delmlsnd}).}  so we will look for solutions
of neutrino oscillations among four species where the mass squared
differences are the three mass scales mentioned above.

In the present case there are six possibilities, and they are
classified into two categories.  One is a case in which two mass
eigenstates have different degenerate mass scales:
\begin{enumerate}
\renewcommand{\labelenumi}{(\roman{enumi})}
\begin{enumerate}
\setcounter{enumi}{1}
\renewcommand{\labelenumii}{(\roman{enumi}\alph{enumii})}
\item
\begin{eqnarray}
m_{1}^2 \simeq m_{2}^2 &\ll& m_{3}^2 \simeq m_{4}^2\nonumber\\
&{\rm with}& (\Delta m_{21}^2, \Delta m_{43}^2)=
(\Delta m_\odot^2,\Delta m_{\rm atm}^2)\nonumber\\
&{\ }&\Delta m_{jk}^2 = \Delta m_{\rm LSND}^2 ~{\rm for}
{}~ j=3,4,k=1,2,
\label{eqn:patternia}
\end{eqnarray}
\item
\begin{eqnarray}
m_{1}^2 \simeq m_{2}^2 &\ll& m_{3}^2 \simeq m_{4}^2\nonumber\\
&{\rm with}& (\Delta m_{21}^2, \Delta m_{43}^2)=
(\Delta m_{\rm atm}^2,\Delta m_\odot^2)\nonumber\\
&{\ }&\Delta m_{jk}^2 = \Delta m_{\rm LSND}^2 ~{\rm for}
{}~ j=3,4,k=1,2.
\label{eqn:patternib}
\end{eqnarray}
\end{enumerate}
\end{enumerate}
\noindent
Another is a case in which three mass eigenstates have degenerate
masses while one eigenstate has a mass far from others:
\begin{enumerate}
\renewcommand{\labelenumi}{(\roman{enumi})}
\begin{enumerate}
\setcounter{enumi}{2}
\renewcommand{\labelenumii}{(\roman{enumi}\alph{enumii})}
\item
\begin{eqnarray}
m_{1}^2 \simeq m_{2}^2 \simeq m_{3}^2 &\ll& m_{4}^2\nonumber\\
&{\rm with}& (\Delta m_{21}^2, \Delta m_{32}^2)=
(\Delta m_\odot^2,\Delta m_{\rm atm}^2)\nonumber\\
&{\ }&\Delta m_{4j}^2 = \Delta m_{\rm LSND}^2 ~{\rm for}~
j=1,2,3,
\label{eqn:patterniia}
\end{eqnarray}
\item
\begin{eqnarray}
m_{1}^2 \simeq m_{2}^2 \simeq m_{3}^2 &\ll& m_{4}^2\nonumber\\
&{\rm with}& (\Delta m_{21}^2, \Delta m_{32}^2)=
(\Delta m_{\rm atm}^2,\Delta m_\odot^2)\nonumber\\
&{\ }&\Delta m_{4j}^2 = \Delta m_{\rm LSND}^2 ~{\rm for}~
j=1,2,3,
\label{eqn:patterniib}
\end{eqnarray}
\item
\begin{eqnarray}
m_{1}^2 &\ll& m_{2}^2 \simeq m_{3}^2 \simeq m_{4}^2\nonumber\\
&{\rm with}& (\Delta m_{32}^2, \Delta m_{43}^2)=
(\Delta m_\odot^2,\Delta m_{\rm atm}^2)\nonumber\\
&{\ }&\Delta m_{j1}^2 = \Delta m_{\rm LSND}^2 ~{\rm for}~
 j=2,3,4,
\label{eqn:patterniic}
\end{eqnarray}
\item
\begin{eqnarray}
m_{1}^2 &\ll& m_{2}^2 \simeq m_{3}^2 \simeq m_{4}^2\nonumber\\
&{\rm with}& (\Delta m_{32}^2, \Delta m_{43}^2)=
(\Delta m_{\rm atm}^2,\Delta m_\odot^2)\nonumber\\
&{\ }&\Delta m_{j1}^2 = \Delta m_{\rm LSND}^2 ~{\rm for}~
 j=2,3,4.
\label{eqn:patterniid}
\end{eqnarray}
\end{enumerate}
\end{enumerate}

The latter four possibilities (ii-a) -- (ii-d) are excluded by reactor
and accelerator experiments, as we will show below.  The case (ib) can
be treated in exactly the same manner as (ia) by changing the labels
1$\leftrightarrow$3, 2$\leftrightarrow$4 of the mass eigenstates, so
in sections 3 and 4 we will discuss only the case (ia) for simplicity.

\section{Reactor and Accelerator experiments}

Let us discuss constraints from reactor and accelerator experiments.
The probability $P(\nu_\alpha\rightarrow\nu_\beta)$ of the transition
$\nu_\alpha\rightarrow\nu_\beta$ in vacuum is given by
\begin{eqnarray}
P(\nu_\alpha\rightarrow\nu_\beta)
&=&-4\sum_{i<j}U_{\alpha i} U_{\alpha j}^\ast
U_{\beta i}^\ast U_{\beta j}
\sin^2\left({\Delta E_{ij}L \over 2}\right)
\label{eqn:a2b}\nonumber\\
&{}&-2\,\sum_{i<j}{\rm Im}\left(U_{\alpha i} U_{\alpha j}^\ast
U_{\beta i}^\ast U_{\beta j}\right)
\sin\left(\Delta E_{ij}L\right)\\
P(\nu_\alpha\rightarrow\nu_\alpha)
&=&1-4\sum_{i<j}|U_{\alpha i}|^2 |U_{\alpha j}|^2
\sin^2\left({\Delta E_{ij}L \over 2}\right),
\label{eqn:a2a}
\end{eqnarray}
where $\alpha=e,\mu,\tau,s$ are the flavor indices,
$\Delta E_{ij}\equiv E_i-E_j=\sqrt{p^2+m_i^2}-\sqrt{p^2+m_j^2}
\simeq (m_i^2-m_j^2)/2E\equiv\Delta m_{ij}^2/2E$ is the difference of
the energy of the two mass eigenstates.  (\ref{eqn:a2b}) and
(\ref{eqn:a2a}) are exact expressions in vacuum, and they are
simplified if we assume the patterns
(\ref{eqn:patternia}) -- (\ref{eqn:patterniid})
with mass hierarchy.

\subsection{Mass Pattern (i)}

Applying the formula (\ref{eqn:a2b}) and (\ref{eqn:a2a}) to the present case
(\ref{eqn:patternia}) with
mass hierarchy, we obtain the following expressions:
\begin{eqnarray}
P(\nu_\alpha\rightarrow\nu_\beta)
&\simeq&4\left\vert \sum_{j=3}^4 U_{\alpha j}U_{\beta j}^\ast
\right\vert^2
\sin^2\left({\Delta E_{31}L \over 2}\right)
\label{eqn:a2bi}\\
P(\nu_\alpha\rightarrow\nu_\alpha)
&\simeq&1-4(U_{\alpha 1}|^2+|U_{\alpha 2}|^2)
(U_{\alpha 3}|^2+|U_{\alpha 4}|^2)
\sin^2\left({\Delta E_{31}L \over 2}\right),
\label{eqn:a2ai}
\end{eqnarray}
where the CP violating terms have been dropped out in
(\ref{eqn:a2bi}).
{}From (\ref{eqn:a2ai}) the
negative result of a disappearance experiment measured at
distances $L_1$ and $L_2$ gives a condition:
\begin{eqnarray}
\epsilon >&{ }& 4|U_{\alpha 1}|^2|U_{\alpha 2}|^2
\epsilon f_\alpha(\Delta m_{21}^2)\nonumber\\
&+&4(U_{\alpha 1}|^2+|U_{\alpha 2}|^2)(U_{\alpha 3}|^2+|U_{\alpha 4}|^2)
\epsilon f_\alpha(\Delta m_{31}^2)\nonumber\\
&+&4|U_{\alpha 3}|^2|U_{\alpha 4}|^2
\epsilon f_\alpha(\Delta m_{43}^2).
\label{eqn:disappear}
\end{eqnarray}
$f_\alpha(\Delta m^2)$ in (\ref{eqn:disappear}) is defined by
\begin{eqnarray}
f_\alpha(\Delta m^2)&\equiv&{1 \over \epsilon}
\left[\left\langle\sin^2\left({\Delta m^2L_1 \over 4E}\right)
\right\rangle_{\alpha\rightarrow\alpha}
-\left\langle\sin^2\left({\Delta m^2L_2 \over 4E}\right)
\right\rangle_{\alpha\rightarrow\alpha}
\right]\nonumber\\
&{ }&\left\{\begin{array}{ll}
=1 / \sin^22\theta_\alpha(\Delta m^2)
\quad&{\rm for}~f_\alpha(\Delta m^2)\ge 1\nonumber\\
\simeq{\rm const.}(\Delta m^2)^2\quad&{\rm for}
{}~\left\langle\sin^2\left({\Delta m^2L_1 \over 4E}\right)
\right\rangle\ll 1,
\end{array}\right.
\label{eqn:f}\\
\end{eqnarray}
where
\begin{eqnarray}
&{ }&\left\langle\sin^2\left({\Delta m^2L \over 4E}\right)
\right\rangle_{\alpha\rightarrow\beta}\nonumber\\
&\equiv&{1 \over
N_{\alpha\beta}(L)}n_T \int_0^\infty dE
\int_0^{q_{\rm max}} dq~\epsilon(q)F_\alpha (E)
{d\sigma_\beta(E,q) \over dq}
\sin^2\left({\Delta m^2L \over 4E}\right),\nonumber\\
\\
N_{\alpha\beta}(L)&\equiv& n_T\int_0^\infty dE
\int_0^{q_{\rm max}} dq~\epsilon(q)F_\alpha (E)
{d\sigma_\beta(E,q) \over dq},
\end{eqnarray}
$F_\alpha (E)$ is the flux of neutrino $\nu_\alpha$ with energy
$E$, $n_T$ is the number of target nucleons, $\epsilon(q)$ is
the detection efficiency function for charged leptons $\ell_\beta$
of energy $q$, $d\sigma_\beta(E,q)/dq$ is the differential cross
section of the interaction $\nu_\beta X\rightarrow\ell_\beta X'$,
and $\epsilon$ stands for the largest fraction of the appearance events
allowed by a given confidence level.

Let us first discuss the reactor experiment \cite{bugey}.
In this case we have $|\Delta m_{21}^2L/4E|\ll 1$, so that
(\ref{eqn:disappear}) becomes
\begin{eqnarray}
{1 \over 4} &>& (|U_{e3}|^2+|U_{e4}|^2)
(1-|U_{e3}|^2-|U_{e4}|^2)f_{\rm Bugey}
\left(\Delta m_{31}^2\right)\nonumber\\
&{ }&+|U_{e3}|^2|U_{e4}|^2f_{\rm Bugey}\left(\Delta m_{43}^2\right),
\label{eqn:bugey1}
\end{eqnarray}
where $f_{\rm Bugey}\left(\Delta m_{31}^2\right)$ stands for
$f_e\left(\Delta m_{31}^2\right)$ in (\ref{eqn:f}) in case
of the Bugey experiment \cite{bugey}.  From the
$(\Delta m^2,\sin^22\theta)$ plot in \cite{bugey}
we have
\begin{eqnarray}
6\times 10^{-2}\lsim &f_{\rm Bugey}\left(\Delta m_{43}^2\right)&
\lsim 25\quad{\rm for}\quad3\times 10^{-3} {\rm eV}^2
\lsim\Delta m_{43}^2\lsim9\times 10^{-2} {\rm eV}^2\nonumber\\
11\lsim &f_{\rm Bugey}\left(\Delta m_{31}^2\right)&
\lsim 25\quad{\rm for}\quad 0.4{\rm eV}^2
\lsim\Delta m_{31}^2\lsim 2.5{\rm eV}^2.
\label{eqn:bugey2}
\end{eqnarray}

Since each term in (\ref{eqn:bugey1}) is semipositive definite,
from (\ref{eqn:bugey1}) and (\ref{eqn:bugey2}) we get
\begin{eqnarray}
|U_{e3}|^2+|U_{e4}|^2 < {1 \over 2}
\left[ 1 -\sqrt{1-1/f_{\rm Bugey}\left(\Delta m_{31}^2\right)}\right]
\label{eqn:bugey3}
\end{eqnarray}
or
\begin{eqnarray}
|U_{e3}|^2+|U_{e4}|^2 > {1 \over 2}\left[1
+\sqrt{1-1/f_{\rm Bugey}\left(\Delta m_{31}^2\right)}\right].
\label{eqn:bugey4}
\end{eqnarray}
On the other hand, we have the following relation between the
probabilities $P^{(2)}\left(\nu_e\rightarrow\nu_e;A(x)\right)$
and $P^{(4)}\left(\nu_e\rightarrow\nu_e;A(x)\right)$ for
the solar neutrinos transitions in the two and four flavor mixings,
respectively:
\begin{eqnarray}
P^{(4)}\left(\nu_e\rightarrow\nu_e;A(x)\right)
&=&P^{(2)}\left(\nu_e\rightarrow\nu_e;
(1-|U_{e3}|^2-|U_{e4}|^2)A(x)\right)\nonumber\\
&\times&\left(1-|U_{e3}|^2-|U_{e4}|^2\right)^2
+|U_{e3}|^2+|U_{e4}|^2,
\label{eqn:prob}
\end{eqnarray}
where $A(x)$ stands for the electron density, and we have averaged
over rapid oscillations $\sin^2(\Delta m_{ij}^2L/4E)$ for all
$(i,j)\ne(2,1)$.  Eq. (\ref{eqn:prob}) can be derived in a way similar
to the case of the relation between $P^{(2)}$ and $P^{(3)}$
\cite{lim}\cite{smirnov}.  To account for the suppression of the
${}^7$Be neutrino among all the solar neutrinos,
$P^{(4)}\left(\nu_e\rightarrow\nu_e;A(x)\right)$ cannot be larger than
${1 \over 2}$ \cite{bgk}, so the second possibility (\ref{eqn:bugey4})
is excluded.  Using the relations $U_{e3}=s_{13}c_{14}e^{i\delta_2}$,
$U_{e4}=s_{14}e^{i\delta_3}$, which can be derived from the expression
(\ref{eqn:vkm}), we have
\begin{eqnarray}
s_{13}^2,~s_{14}^2\lsim 1/f_{\rm Bugey}\left(\Delta m_{31}^2\right)
\lsim{1 \over 40}\quad{\rm for}\quad 0.24
{\rm eV}^2\lsim\Delta m_{31}^2\lsim 2.5{\rm eV}^2
\end{eqnarray}
or
\begin{eqnarray}
\theta_{13},~\theta_{14}\lsim 9^\circ.
\label{eqn:s13s14}
\end{eqnarray}

Next let us consider the disappearance experiment of
$\nu_\mu\rightarrow\nu_\mu$ \cite{CDHSW}.  In this case
we have $|\Delta m_{21}^2L/4E|, |\Delta m_{43}^2L/4E|\ll 1$,
so that
\begin{eqnarray}
{1 \over 4} > (|U_{\mu3}|^2+|U_{\mu4}|^2)
(1-|U_{\mu3}|^2-|U_{\mu4}|^2)
{}~f_{\rm CDHSW}\left(\Delta m_{31}^2\right)
\end{eqnarray}
where $f_{\rm CDHSW}\left(\Delta m_{31}^2\right)$ stands for
$f_\mu\left(\Delta m_{31}^2\right)$ in (\ref{eqn:f})
for the CDHSW data \cite{CDHSW}.  The allowed value for
$f_{\rm CDHSW}\left(\Delta m_{31}^2\right)$ is given by
\begin{eqnarray}
1\lsim f_{\rm CDHSW}\left(\Delta m_{31}^2\right)
\lsim 50\quad{\rm for}\quad
0.24{\rm eV}^2\le \Delta m_{31}^2\le 2.5{\rm eV}^2,
\end{eqnarray}
and therefore we obtain
\begin{eqnarray}
|U_{\mu3}|^2+|U_{\mu4}|^2 < {1 \over 2}\left[1
-\sqrt{1-1/f_{\rm CDHSW}\left(\Delta m_{31}^2\right)}\right]
\label{eqn:cdhsw1}
\end{eqnarray}
or
\begin{eqnarray}
|U_{\mu3}|^2+|U_{\mu4}|^2 > {1 \over 2}\left[1
+\sqrt{1-1/f_{\rm CDHSW}\left(\Delta m_{31}^2\right)}\right].
\label{eqn:cdhsw2}
\end{eqnarray}

Furthermore, we have the constraints from the atmospheric neutrino
data.  If we consider the probability $P(\nu_\mu\rightarrow\nu_\mu)$
for the multi-GeV data by the Kamiokande group, then we get
\begin{eqnarray}
P(\nu_\mu\rightarrow\nu_\mu)&=&
1-4|U_{\mu3}|^2|U_{\mu4}|^2\sin^2\left(
{\Delta m_{43}^2L \over 4E}\right)\nonumber\\
&{ }&-2(|U_{\mu3}|^2+|U_{\mu4}|^2)
(1-|U_{\mu3}|^2-|U_{\mu4}|^2),
\label{eqn:kamioka}
\end{eqnarray}
where we have used $|\Delta m_{21}^2L/4E|\ll 1$.  To have the
zenith angle dependence of the multi-GeV data,
it is necessary for $P(\nu_\mu\rightarrow\nu_\mu)$
to deviate from unity to some extent.
Using the constraints $s_{13}^2, s_{14}^2\lsim 1/40$ in
(\ref{eqn:s13s14}) and the relations
\begin{eqnarray}
U_{\mu1}&=&-c_{12}c_{23}c_{24}s_{13}e^{-i\delta_2}
+c_{24}s_{12}s_{23}e^{i\delta_1}
-c_{12}c_{13}s_{14}s_{24}e^{-i\delta_3}\\
U_{\mu2}&=&-c_{23}c_{24}s_{12}s_{13}e^{-i\delta_2}
-c_{12}c_{24}s_{23}e^{i\delta_1}
-c_{13}s_{12}s_{14}s_{24}e^{-i\delta_3},
\end{eqnarray}
we have
\begin{eqnarray}
|U_{\mu1}|^2+|U_{\mu2}|^2&\simeq&s_{23}^2c_{24}^2\nonumber\\
&<& {1 \over 2}\left[1
-\sqrt{1-1/f_{\rm CDHSW}\left(\Delta m_{31}^2\right)}\right].
\label{eqn:mu1mu2}
\end{eqnarray}
Because of the condition (\ref{eqn:s13s14}), the probability
$P(\nu_e\rightarrow\nu_e)$ is close to unity for the mass
squared difference $\Delta m^2\sim\Delta m_{\rm atm}^2$, so the
only possible source for the atmospheric neutrino anomaly
in our scenario is deviation of the probability
$P(\nu_\mu\rightarrow\nu_\mu)$ from unity.
Using the technique in \cite{y},
we find that the region of the parameters $|U_{\mu3}|^2$ and
$|U_{\mu4}|^2$ allowed for the multi-GeV atmospheric
neutrinos data of Kamiokande at 90\% confidence level
is
\begin{eqnarray}
0\le\theta_{23}\lsim 50^\circ,~25^\circ\lsim\theta_{24}\lsim 55^\circ,
\label{eqn:s23s24}
\end{eqnarray}
where we have used the relations
\begin{eqnarray}
U_{\mu3}&=&c_{13}c_{23}c_{24}
-s_{13}s_{14}s_{24}e^{i(\delta_2-\delta_3)}\\
U_{\mu4}&=&c_{14}s_{24}.
\end{eqnarray}
{}From (\ref{eqn:s23s24}) we have $s_{23}^2c_{24}^2\lsim 0.48$, and
hence the possibility (\ref{eqn:cdhsw1}) is excluded.

Let us now consider the LSND data \cite{lsnd}.  In this case
we have
\begin{eqnarray}
|\Delta m_{21}^2L/4E|, |\Delta m_{43}^2L/4E|\ll 1
\end{eqnarray}
so we obtain
\begin{eqnarray}
\epsilon = 4\epsilon |U_{e3}U_{\mu3}^\ast+U_{e4}U_{\mu4}^\ast|^2
f_{\rm LSND}\left(\Delta m_{31}^2\right),
\label{eqn:lsnd1}
\end{eqnarray}
with
\begin{eqnarray}
f_{\rm LSND}\left(\Delta m_{31}^2\right)
&\equiv&{1 \over \epsilon}
\left\langle\sin^2\left({\Delta m^2L \over 4E}\right)
\right\rangle_{\mu\rightarrow e}
\\
&{ }&\left\{\begin{array}{ll}
=1 / \sin^22\theta_{\rm LSND}(\Delta m^2)
\quad&{\rm for}~\Delta m^2 > 0.1{\rm eV}^2\nonumber\\
\simeq{\rm const.}(\Delta m^2)^2\quad&{\rm for}
{}~\Delta m^2 < 0.1{\rm eV}^2,
\end{array}\right.
\end{eqnarray}
where $\sin^22\theta_{\rm LSND}(\Delta m^2)$ is the value in the
allowed region in the $(\Delta m^2,\sin^22\theta)$ plot
in \cite{lsnd}, and $25\lsim f_{\rm LSND}(\Delta m_{31}^2)
\lsim 580$.  Using the conditions on $s_{13},~s_{14},
{}~s_{23}$, $s_{24}$, we obtain
\begin{eqnarray}
1\simeq 4
\left\vert{1 \over \sqrt{2}}\left(s_{13}+s_{14}\right)\right\vert^2
f_{\rm LSND}\left(\Delta m_{31}^2\right)
\end{eqnarray}
and hence
\begin{eqnarray}
9^\circ\gsim\max(\theta_{13},\theta_{14})\gsim 0.8^\circ,
\end{eqnarray}
where we have also shown the upper bound (\ref{eqn:s13s14}).

\subsection{Mass Pattern (ii)}

We now show that the mass patterns (\ref{eqn:patterniia})
-- (\ref{eqn:patterniid}) are excluded by reactor and
accelerator experiments.  The arguments against the four cases are
exactly the same, so for simplicity
we will consider the case (\ref{eqn:patterniia}),
where $\Delta m_{\rm 4j}^2\simeq\Delta m_{\rm LSND}^2$ for $j=1,2,3$.
If the flight length $L$ of neutrinos in the experiment satisfies
$\Delta E_{41} L\sim {\cal O}(1)$, then
the oscillation probabilities are given by
\begin{eqnarray}
P(\nu_\alpha\rightarrow\nu_\beta)
&\simeq&4\vert U_{\alpha 4}\vert^2\vert U_{\beta 4}\vert^2
\sin^2\left({\Delta E_{41}L \over 2}\right)\\
P(\nu_\alpha\rightarrow\nu_\alpha)
&\simeq&1-4\vert U_{\alpha 4}\vert^2
\left(1-\vert U_{\alpha 4}\vert^2\right)
\sin^2\left({\Delta E_{41}L \over 2}\right),
\end{eqnarray}
where CP violating terms have been dropped in
$P(\nu_\alpha\rightarrow\nu_\beta)$.
Exactly with the same arguments as in the previous subsection,
the data by Bugey in this case implies
\begin{eqnarray}
|U_{e4}|^2 < {1 \over 2}
\left[ 1 -\sqrt{1-1/f_{\rm Bugey}\left(\Delta m_{41}^2\right)}\right],
\label{eqn:bugeyii}
\end{eqnarray}
the CDHSW data leads to
\begin{eqnarray}
|U_{\mu4}|^2 < {1 \over 2}\left[1
-\sqrt{1-1/f_{\rm CDHSW}\left(\Delta m_{41}^2\right)}\right],
\label{eqn:cdhswii}
\end{eqnarray}
and the LSND data to
\begin{eqnarray}
1 = 4 |U_{e4}|^2|U_{\mu4}|^2
f_{\rm LSND}\left(\Delta m_{41}^2\right).
\label{eqn:lsndii}
\end{eqnarray}
So $\Delta m_{41}^2$ has to satisfy the following constraint:
\begin{eqnarray}
&{\ }&4 |U_{e4}|^2|U_{\mu4}|^2\nonumber\\
&=&{1 \over f_{\rm LSND}\left(\Delta m_{41}^2\right)}\nonumber\\
&<&\left[ 1 -\sqrt{1-{1 \over f_{\rm Bugey}
\left(\Delta m_{41}^2\right)}}\right]
\left[1 -\sqrt{1-{1 \over f_{\rm CDHSW}
\left(\Delta m_{41}^2\right)}}\right].
\label{eqn:delm41}
\end{eqnarray}
It turns out that the condition (\ref{eqn:delm41}) is not satisfied
for any $\Delta m_{41}^2$ in the entire region
$0.27{\rm eV^2}\lsim\Delta m_{41}^2\lsim 2.3{\rm eV}^2$.
Hence the mass patterns (\ref{eqn:patterniia}) -- (\ref{eqn:patterniid})
are excluded by reactor and accelerator experiments.

\section{Big Bang Nucleosynthesis}

Let us now discuss the constraints by cosmology.  It has been shown
\cite{bbn1}\cite{bbn2}\cite{bbn3} that the transitions
$\nu_\alpha\rightarrow\nu_s~(\alpha=e,\mu,\tau)$ have to be suppressed
strongly for big bang nucleosynthesis to be consistent with the
standard scenario, if the number $N_\nu$ of light neutrinos is less
than four.\footnote{For recent discussions on $N_\nu$, see
\cite{bbn4}\cite{bbn5} and references therein.}  Here we follow the
argument of \cite{bbn1}\cite{bbn2} and give a rough estimate on the
allowed region of the mixing angles without detailed numerical
calculations.  Except in section 7, we will assume $N_\nu < 4 $ in the
following sections.  We will show in section 7 that if $N_\nu \ge 4 $
we have much weaker constraints by big bang nucleosynthesis compared
to the case for $N_\nu < 4 $.

To get a rough idea on the magnitude of the effective
mixing angle of neutrino oscillations, let us consider
the temperature dependence of the difference $\Delta E$
of the kinetic term and the potential term $V$
\cite{bbn1}\cite{bbn2}\cite{bbn3}.
\begin{eqnarray}
\Delta E &=& {\Delta m^2 \over 6.3T}
\label{eqn:dele}\\
V&=&-{\sqrt{2}(7/90)^2\pi^6(2+\cos^2\theta_W)
G_F \over  \zeta(3) m_W^2}T^5.
\label{eqn:v}
\end{eqnarray}
{}From (\ref{eqn:dele}) and (\ref{eqn:v}) we have
\begin{eqnarray}
{V \over \Delta E}\simeq
-\left({T \over 13 {\rm MeV}}\right)^6
\left({\Delta m^2 \over {\rm eV}^2}\right)^{-1}.
\end{eqnarray}
In case of the two flavor  neutrino oscillation
with a mixing angle $\theta$,
the interaction rate of sterile neutrinos
through neutrino oscillations is given by
\begin{eqnarray}
\left({\Gamma_{\nu_s} \over H}\right)_{\rm 2-flavor}=\sin^22\theta_M
{}~\sin^2\left({\cal E}\ell_{\rm coll}\right){\Gamma_\nu \over H},
\end{eqnarray}
where $\theta_M$ is the effective mixing angle in the presence of
interactions defined by
\begin{eqnarray}
\sin^22\theta_M\equiv
{\sin^22\theta \over (\cos2\theta-V/\Delta E)^2+\sin^22\theta},
\label{eqn:thetam}
\end{eqnarray}
\begin{eqnarray}
H\equiv \left[{4\pi^3 \over 45} {g_\ast(T) \over m_{\rm pl}^2}
T^4\right]^{1/2}
\end{eqnarray}
is the Hubble parameter, $g_\ast(T)=10.75$ is the
effective degrees of freedom of particles for $m_e\lsim T \lsim m_\mu$,
\begin{eqnarray}
{\cal E}\equiv{1 \over 2}
\left(\Delta E^2 +V^2 -2V\Delta E\cos2\theta\right)^{1/2}
\end{eqnarray}
is the eigenvalue of the mass matrix
\begin{eqnarray}
e^{i\sigma_2\theta}{\rm diag}\left(E_1,E_2\right)e^{-i\sigma_2\theta}
+{\rm diag}\left(V,0\right),
\end{eqnarray}
\begin{eqnarray}
\ell_{\rm coll}\equiv 1/\Gamma_{\nu}=1/C(T)G_F^2T^5
\end{eqnarray}
is the collision length of neutrino, $\Gamma_{\nu}$ is an interaction rate
for active neutrinos,
$C(T)\simeq 0.5$ for $m_e\lsim T \lsim m_\mu$, and
\begin{eqnarray}
{\Gamma_\nu \over H}\simeq \left({T \over 1.9 {\rm MeV}}\right)^3.
\end{eqnarray}
Notice that there is no enhancement of oscillations due to the MSW
mechanism \cite{msw} in the present case, because $\theta_M<\theta$
follows from $V/\Delta E<0$ in (\ref{eqn:thetam}).  Since we are
interested in the mass squared differences $\Delta m_{21}^2\sim {\cal
O}(10^{-5}{\rm eV}^2)$ or ${\cal O}(10^{-10}{\rm eV}^2)$, $\Delta
m_{43}^2\sim {\cal O}(10^{-2}{\rm eV}^2)$, $\Delta m_{31}^2\sim {\cal
O}(1{\rm eV}^2)$, we have to consider situations where $\Delta
E_{jk}=\Delta m_{jk}^2/6.3T$ becomes comparable to the absolute
value $|V|$ of the potential,
and this condition is satisfied for the critical temperature $T\sim
2$MeV (or 0.3MeV), 8MeV, 15MeV, respectively.  Off these regions of
$T$, the problem becomes simpler.  If $T \gg 25$MeV, $|V / \Delta E|$
becomes very large and the effective mixing angle is small, so that we
have
\begin{eqnarray}
\left({\Gamma_{\nu_s} \over H}\right)_{\rm 2-flavor}&\le&
\sin^22\theta_M~{\Gamma_\nu \over H}\nonumber\\
&\sim& \sin^22\theta~\left({\Delta E \over V} \right)^2
\left({T \over 1.9 {\rm MeV}}\right)^3\nonumber\\
&\sim& \sin^22\theta~\left({T \over 25 {\rm MeV}}\right)^{-9}
\left({\Delta m^2 \over {\rm eV}^2}\right)^2\ll 1\nonumber\\
&{\ }&\qquad{\rm for}~\gg 25{\rm MeV}.
\end{eqnarray}
On the other hand, if $T \ll 1$MeV, $|V / \Delta E|$
becomes very small and this case is reduced to the
oscillations in vacuum:
\begin{eqnarray}
\left({\Gamma_{\nu_s} \over H}\right)_{\rm 2-flavor}
&\lsim&\sin^22\theta~\left({T \over 1.9 {\rm MeV}}\right)^3
\ll 1\nonumber\\
&{\ }&\qquad{\rm for}~\ll 1{\rm MeV}.
\end{eqnarray}
In the above extreme cases there is little transition
$\nu_s\rightarrow\nu_\alpha$ ($\alpha=e,\mu,\tau$) and
we have no problem with big bang nucleosynthesis.
So in the following discussions we will discuss only the region of
temperature $m_e\lsim T \lsim m_\mu$.  Since $\Delta E \ell_{\rm coll}/2
\sim 250|\Delta E / V|$ for any region of $T$, the term
$\sin^2\left( {\cal E} \ell_{\rm coll}\right)$ in the formula of
neutrino oscillations can be put to 1/2 after averaging over
rapid oscillations, as long as the temperature under consideration
satisfies $|\Delta E / V| \sim {\cal O} (1)$.

To analyze oscillations among four species of neutrinos,
it is necessary to diagonalize the mass matrix
\begin{eqnarray}
{\cal M}\equiv U{\rm diag}\left(E_1, E_2, E_3, E_4\right)U^{-1}
+{\rm diag}\left(V,cV,cV,0\right),
\label{eqn:massmatrix}
\end{eqnarray}
where we have taken into account the fact that $\nu_e$ has both
charged and neutral current interactions while $\nu_\mu$ and $\nu_\tau$
have only neutral ones for $m_e\lsim t\lsim m_\mu$, and
\begin{eqnarray}
c\equiv \cos^2\theta_W/(2+\cos^2\theta_W)\simeq 0.28
\end{eqnarray}
is the ratio of neutral current interactions to the total
contribution.  Here we consider three cases where the potential term
becomes comparable to the energy difference $|V|\simeq\Delta
E_{jk}=\Delta m_{jk}^2/6.3T$ for $(j,k)=(2,1),(4,3),(3,1)$.  This
corresponds to the critical temperature $T\simeq$ 2MeV
(or 0.3MeV), 8MeV, 15MeV, respectively.
{}From the constraint
(\ref{eqn:s13s14}) by reactor experiments we know that
$U_{e3}$ and $U_{e4}$ are small, so we take
$U_{e3}=U_{e4}=0$ in what follows for simplicity.
Since we assume $N_\nu < 4 $,
sterile neutrinos should never have been in thermal
equilibrium, and we demand that the interaction rate $\Gamma_{\nu_s}$
of sterile neutrinos
through neutrino oscillations be smaller than the Hubble parameter $H$:
\begin{eqnarray}
\Gamma_{\nu_s}&=&\sum_{\alpha=e,\mu,\tau}
P(\nu_s\rightarrow\nu_\alpha)\Gamma_{\nu_\alpha}\nonumber\\
&=&\left[ P(\nu_s\rightarrow\nu_e)
+B\left\{ P(\nu_s\rightarrow\nu_\mu)
+P(\nu_s\rightarrow\nu_\tau)\right\}
\right]
\Gamma_{\nu}\nonumber\\
&<& H,
\label{eqn:gammanus}
\end{eqnarray}
where
\begin{eqnarray}
B\equiv
\left({1-4\sin^2\theta_W+8\sin^4\theta_W \over
1+4\sin^2\theta_W +8\sin^4\theta_W}\right)\simeq 0.21.
\end{eqnarray}
In (\ref{eqn:gammanus})
both charged and neutral current interactions are
taken into account for $\nu_e$, while only neutral ones are
included for $\nu_\mu$ and $\nu_\tau$.

Let us first consider the situation where $|V|\simeq \Delta E_{21}$.
For the vacuum oscillation solution, we can give the same argument
and arrive at the same conclusion as the MSW ones.
In what follows, therefore, we will discuss for simplicity the case of
the MSW solutions, which imply a critical temperature $T\simeq 2$MeV.
Now if $T\simeq 2$MeV,
the mass matrix (\ref{eqn:massmatrix}) is diagonalized as
\begin{eqnarray}
{\cal M}\simeq UU_M\left[(E_1+cV){\bf 1}_4
+{\rm diag }\left(\lambda_1,\lambda_2,
\Delta E_{31},\Delta E_{41}\right)\right]U_M^{-1}U^{-1},
\end{eqnarray}
where
\begin{eqnarray}
\lambda_1,\lambda_2&=&\Delta E_{21}\mp{1 \over 2}
\left[\Delta E_{21}^2
-2\Delta E_{21}(1-cA) V\cos2\theta_\odot
+(1-cA) V^2\right]^{1/2},\nonumber\\
\\
U_M&\equiv&\left(
\begin{array}{cc}
e^{i\sigma_2\theta_{12}^M} & {\bf 0}\\
{\bf 0} & {\bf 1}_2\\
\end{array}\right),\\
A&\equiv&\vert U_{s3}\vert^2+\vert U_{s4}\vert^2
\end{eqnarray}
with
\begin{eqnarray}
\tan2\theta_{12}^M\equiv {(1-cA) V\sin2\theta_\odot
\over \Delta E_{21}-(1-cA) V\cos2\theta_\odot}.
\end{eqnarray}
In this case difference of any two eigenvalues of ${\cal M}$
is large compared to the collision length $\ell_{\rm coll}$,
so the factor $\sin^2\left( {\cal E} \ell_{\rm coll}\right)$ can be
put to $1/2$, and the probabilities are given by
\begin{eqnarray}
P(\nu_s\rightarrow\nu_e)&\simeq&{1-A \over 2}
\sin^22(\theta_{12}^M+\theta_\odot)
\label{eqn:nusnue}\\
P(\nu_s\rightarrow\nu_\mu)&+&P(\nu_s\rightarrow\nu_\tau)\nonumber\\
&\simeq&A(1-A)\left\{2-{1 \over 2}\sin^22(\theta_{12}^M
+\theta_\odot)\right\}
+{A^2 \over 2}\sin^22\phi,
\end{eqnarray}
where
\begin{eqnarray}
\sin^22(\theta_{12}^M+\theta_\odot)
&\equiv&{\sin^22\theta_\odot \over
1-2(1-cA) V \cos2\theta_\odot/\Delta E_{21}
+(1-cA)^2 V^2 /\Delta E_{21}^2},\nonumber\\
\label{eqn:th12m}
\end{eqnarray}
and we have parametrized the matrix elements as
\begin{eqnarray}
\left(
\begin{array}{c}
U_{s3}\\
U_{s4}\\
\end{array}\right)&\equiv& A^{1/2}e^{i(\chi+\sigma_3\psi)/2-i\sigma_2\phi}
\left(
\begin{array}{c}
1\\
0\\
\end{array}\right)
\end{eqnarray}
with
\begin{eqnarray}
\chi&\equiv&{\rm arg}(U_{s3}U_{s4}),\\
\psi&\equiv&{\rm arg}(U_{s3}/U_{s4}),\\
\phi&\equiv&{\rm tan}^{-1}\vert U_{s4}/U_{s3}\vert.
\end{eqnarray}
Thus we have the interaction rate of sterile neutrinos
\begin{eqnarray}
\Gamma_{\nu_s}&=&\left[
P(\nu_s\rightarrow\nu_e)+B
\left\{P(\nu_s\rightarrow\nu_\mu)+P(\nu_s\rightarrow\nu_\tau)\right\}
\right]\Gamma_\nu\nonumber\\
&\simeq&\left[B\{2A(1-A)+{A^2 \over 2}\sin^22\phi\}
+\{{1-A \over 2}-{B \over 2}A(1-A)\}\sin^2
2(\theta_{12}^M+\theta_\odot)\right]\Gamma_\nu\nonumber\\
&{\ }&\qquad{\rm for}~T\sim 2{\rm MeV}.
\label{eqn:gammanus1}
\end{eqnarray}
If $T\simeq 8{\rm MeV}$ (i.e., $|V|\simeq \Delta E_{43}$),
the mass matrix (\ref{eqn:massmatrix}) is diagonalized as
\begin{eqnarray}
{\cal M}\simeq UU_M\left[(E_3+cV){\bf 1}_4+{\rm diag }\left(-\Delta E_{31},
-\Delta E_{32},\lambda_3,\lambda_4
\right)\right]U_M^{-1}U^{-1},
\end{eqnarray}
where
\begin{eqnarray}
\lambda_3,\lambda_4&=&\Delta E_{43}\mp{1 \over 2}
\left[\Delta E_{43}^2
+2\Delta E_{43} cAV\cos2\theta_\odot
+c^2A^2V^2\right]^{1/2},\\
U_M&\equiv&\left(
\begin{array}{cc}
{\bf 1}_2 & {\bf 0}\\
{\bf 0} & e^{i\sigma_2\theta_{34}^M}\\
\end{array}\right),
\end{eqnarray}
with
\begin{eqnarray}
\tan2\theta_{34}^M\equiv { cAV\sin2\phi
\over \Delta E_{34}+ cAV\cos2\phi}.
\end{eqnarray}
All differences of two eigenvalues of ${\cal M}$ except $\Delta E_{21}$
are large compared to the collision length $\ell_{\rm coll}$,
and we put all the factors $\sin^2\left( {\cal E} \ell_{\rm coll}\right)$
to $1/2$ but set
$\sin^2\left(\Delta E_{21} \ell_{\rm coll}\right)$ to zero.
The probabilities are given by
\begin{eqnarray}
P(\nu_s\rightarrow\nu_e)&\simeq&0
\\
P(\nu_s\rightarrow\nu_\mu)&+&P(\nu_s\rightarrow\nu_\tau)\nonumber\\
&\simeq&2A(1-A)+{A^2 \over 2}\sin^22(\phi-\theta_{34}^M),
\end{eqnarray}
where
\begin{eqnarray}
\sin^22(\phi-\theta_{34}^M)
&\equiv&{\sin^22\phi \over
1-2 cAV \cos2\phi/\Delta E_{43}
+ c^2A^2V^2 /\Delta E_{43}^2}.
\end{eqnarray}
The interaction rate of sterile neutrinos is given by
\begin{eqnarray}
\Gamma_{\nu_s}&\simeq&
B\left[2A(1-A)+{A^2 \over 2}\sin^22(\phi-\theta_{34}^M)
\right]\Gamma_\nu.\nonumber\\
&{\ }&\qquad{\rm for}~T\sim 8{\rm MeV}
\label{eqn:gammanus2}
\end{eqnarray}
If $T\simeq 15{\rm MeV}$ (i.e., $|V|\simeq \Delta E_{31}$),
the mass matrix (\ref{eqn:massmatrix}) is diagonalized as
\begin{eqnarray}
{\cal M}\simeq UU_M\left[E_1{\bf 1}_4+{\rm diag }\left(V,
-\Delta E_{31},\lambda_5,\lambda_6
\right)\right]U_M^{-1}U^{-1},
\end{eqnarray}
where
\begin{eqnarray}
\lambda_5,\lambda_6&=&{-\Delta E_{31}+cV \over 2}\pm{1 \over 2}
\left[(\Delta E_{31}^2-cV)^2
+4cAV\Delta E_{31}\right]^{1/2},\\
U_M&\equiv&\left(
\begin{array}{cccc}
\begin{array}{c}
U_{e1}\\
U_{e2}
\end{array}&
\begin{array}{c}
0\\
0
\end{array}&\alpha_+\left(
\begin{array}{c}
U_{s1}^\ast\\
U_{s2}^\ast
\end{array}\right)&\alpha_-\left(
\begin{array}{c}
U_{s1}^\ast\\
U_{s2}^\ast
\end{array}\right)\\
\begin{array}{c}
0\\
0
\end{array}&A^{{}^{-{1 \over 2}}}\left(
\begin{array}{c}
U_{s4}\\
-U_{s3}
\end{array}\right)&\beta_+\left(
\begin{array}{c}
U_{s3}^\ast\\
U_{s4}^\ast
\end{array}\right)&\alpha_-\left(
\begin{array}{c}
U_{s3}^\ast\\
U_{s4}^\ast
\end{array}\right)
\end{array}\right).
\label{eqn:um}
\end{eqnarray}
In (\ref{eqn:um}) $\alpha_\pm, \beta_\pm$ satisfy the normalization
conditions
\begin{eqnarray}
A\vert\beta\vert^2+(1-A)\vert\alpha\vert^2=1,
\end{eqnarray}
and $x_\pm\equiv\beta_\pm/\alpha_\pm$ are the two roots of
a quadratic equation
\begin{eqnarray}
Ax^2+(1-2A+\Delta E_{31}/cV)x-(1-A)=0.
\end{eqnarray}
{}From (\ref{eqn:um}) it turns out that $(UU_M)_{e1}=1$, $(UU_M)_{ej}=0$
for $j=2,3,4$, $(UU_M)_{sj}=0$ for $j=3,4$, and
the only combination which appears
in the formula of probability with non-vanishing coefficients
$(UU_M)_{sj}(UU_M)_{sk}^\ast(UU_M)_{ej}^\ast$ $(UU_M)_{ek}$
is $(j,k)=(3,4)$, and the difference of these two eigenvalues of ${\cal M}$
is large compared to the collision length $\ell_{\rm coll}$.
Hence we again put all the factors
$\sin^2\left( {\cal E} \ell_{\rm coll}\right)$
to $1/2$, and we obtain the probabilities
\begin{eqnarray}
P(\nu_s\rightarrow\nu_e)&\simeq&0
\\
P(\nu_s\rightarrow\nu_\mu)&+&P(\nu_s\rightarrow\nu_\tau)\nonumber\\
&\simeq&2\vert\alpha_+\vert^2\vert\alpha_-\vert^2
\vert 1-A+x_+A\vert^2\vert1-A+x_-A\vert^2\\
&=&{2A(1-A) \over 1+2(1-2A)cV/\Delta E_{31}+c^2V^2/\Delta E_{31}^2}.
\end{eqnarray}
The interaction rate of sterile neutrinos is then given by
\begin{eqnarray}
\Gamma_{\nu_s}&\simeq&
{2c^2A(1-A) \over 1+2(1-2A)cV/\Delta E_{31}+c^2V^2/\Delta E_{31}^2}
\Gamma_\nu.\nonumber\\
&{\ }&\qquad{\rm for}~T\sim 15{\rm MeV}
\label{eqn:gammanus3}
\end{eqnarray}
Combining the results (\ref{eqn:gammanus1}), (\ref{eqn:gammanus2})
and (\ref{eqn:gammanus3}) for three cases we have the following
formula
\begin{eqnarray}
{\Gamma_{\nu_s} \over H}&\simeq&\left({T \over 1.8{\rm MeV}}\right)^3
{}~\left[
{2BA(1-A) \over 1+2(1-2A)cV/\Delta E_{31}
+c^2V^2/\Delta E_{31}^2}\right.\nonumber\\
&+&{BA^2\sin^22\phi/2 \over 1-2 cAV \cos2\phi/\Delta E_{43}
+ c^2A^2V^2 /\Delta E_{43}^2}\nonumber\\
&+&\left.{(1-A)(1-BA)\sin^22\theta_\odot / 2 \over
1-2(1-cA) V \cos2\theta_\odot/\Delta E_{21}
+(1-cA)^2 V^2 /\Delta E_{21}^2}\right],
\label{eqn:gammanus4}
\end{eqnarray}
which approximately holds for any temperature $m_e\lsim T\lsim m_\mu$
\footnote{Notice
that results (\ref{eqn:gammanus1}), (\ref{eqn:gammanus2}),
(\ref{eqn:gammanus3}) are reproduced for each case
in (\ref{eqn:gammanus4}):
{}~the first term $\rightarrow 2c^2A(1-A)$,
the second $\rightarrow c^2A^2\sin^22\phi/2$ for $T\sim 2$MeV
where $V/\Delta E_{31},V/\Delta E_{43}\rightarrow 0$,
the first term $\rightarrow 2c^2A(1-A)$,
the third $\sim(T/2{\rm MeV})^{-12}\rightarrow 0$ for $T\sim 8$MeV
where $V/\Delta E_{31}\rightarrow 0,|V/\Delta E_{21}|\rightarrow$ large,
the second $\sim(T/8{\rm MeV})^{-12}\rightarrow 0$,
the third $\sim(T/2{\rm MeV})^{-12}\rightarrow 0$ for $T\sim 15$MeV
where $|V/\Delta E_{31}|,|V/\Delta E_{43}|\rightarrow$ large.}.
As in \cite{bbn2} we look for the extremum of the expression
$\Gamma_{\nu_s}/H$ with respect to $T$.
The extrema of the three terms in (\ref{eqn:gammanus4})
are attained at different values of $T$ which are far
apart, so we can discuss the extremum of the total quantity
(\ref{eqn:gammanus4}) by examining each term separately.
Each term in
(\ref{eqn:gammanus4}) has the following maximum value
\begin{eqnarray}
\displaystyle\max_T~({\rm 1st~term})&\simeq&\left({15 \over 1.8}\right)^3
\left({\Delta m_{31}^2 \over {\rm eV^2}}\right)^{1/2}
{3B \over 2c^{1/2}}\nonumber\\
&\times&{A(1-A)\left[\left[1/3+((1-2A)/3)^2
\right]^{1/2}-(1-2A)/3\right]^{1/2} \over
1-(1-2A)^2/3+(1-2A)\left[1/3+((1-2A)/3)^2\right]^{1/2}}\nonumber\\
\label{eqn:max1}\\
\displaystyle\max_T~({\rm 2nd~term})&\simeq&\left({7.7 \over 1.8}\right)^3
\left({\Delta m_{43}^2 \over 2\times10^{-2}{\rm eV^2}}\right)^{1/2}
{3A^{3/2}B \over 8c^{1/2}}
\nonumber\\
&\times&{\sin^22\phi\left[\left[1/3+(\cos2\phi/3)^2
\right]^{1/2}+\cos2\phi/3\right]^{1/2} \over
1+\cos^22\phi/3+\cos2\phi\left[1/3+(\cos2\phi/3)^2\right]^{1/2}}
\label{eqn:max2}\\
\displaystyle\max_T~({\rm 3rd~term})&\simeq&\left({1.9 \over 1.8}\right)^3
\left({\Delta m_{21}^2 \over 10^{-5}{\rm eV^2}}\right)^{1/2}
{3(1-A)(1-BA) \over 8(1-cA)^{1/2}}\nonumber\\
&\times&{\sin^22\theta_\odot\left[\left[1/3+(\cos2\theta_\odot/3)^2
\right]^{1/2}-\cos2\theta_\odot/3\right]^{1/2} \over
1-\cos^22\theta_\odot/3+\cos2\theta_\odot
\left[1/3+(\cos2\theta_\odot/3)^2\right]^{1/2}}.
\label{eqn:max3}
\end{eqnarray}
The term in the second line in
(\ref{eqn:max1}) is a monotonically increasing function in A,
so the first term in (\ref{eqn:gammanus4})
gives a necessary condition
\begin{eqnarray}
A&\lsim&{8 \over 3\sqrt{3}}\left({1.8 \over 15}\right)^3
{c^{1/2} \over B}
\left({\Delta m_{31}^2 \over {\rm eV^2}}\right)^{-1/2}\lsim
1.3\times 10^{-2},
\label{eqn:a}
\end{eqnarray}
where we have used the property that the term in the second line
in (\ref{eqn:max1}) is approximated as $\sqrt{3}A/4$ as $A\rightarrow0$ and
the lower bound $\Delta m_{31}^2\gsim 0.27$eV$^2$ in
(\ref{eqn:delmlsnd}).
It is straightforward to see that
the term in the second line in
(\ref{eqn:max2}) is less 4/3
for any value of $\cos2\phi$, and it follows from (\ref{eqn:a})
\begin{eqnarray}
\displaystyle\max_T~({\rm 2nd~term})&\lsim&2\times10^{-2}
\left({\Delta m_{43}^2 \over 2\times10^{-2}{\rm eV^2}}\right)^{1/2}
\ll 1,
\end{eqnarray}
where we have used the combined Kamiokande results of sub-GeV
\cite{kamioka1} and multi-GeV \cite{kamioka2} atmospheric neutrinos,
which suggest $5\times 10^{-3}$eV$^2\lsim\Delta m_{43}^2 \lsim
3\times10^{-2}$eV$^2$.  So the contribution coming from $T\sim 8{\rm
MeV}$ never brings $\nu_s$ into thermal equilibrium, as long as
$\max_{{}_T}({\rm 1st~term})<1$ in (\ref{eqn:max1}) is satisfied.  The
term in the second line in (\ref{eqn:max3}) is approximately less than
0.54 for $\sin^22\theta_\odot\le 0.9$ which is satisfied by both the
MSW solutions.  Thus we have
\begin{eqnarray}
\displaystyle\max_T~({\rm 3rd~term})&\lsim&0.24
\left({\Delta m_{21}^2 \over 10^{-5}{\rm eV^2}}\right)^{1/2},
\end{eqnarray}
so $\nu_s$ is not brought into thermal equilibrium in case
of the MSW solutions, since $\Delta m_{21}^2\lsim 9\times
10^{-5}{\rm eV^2}$ is satisfied by both the MSW solutions.
Exactly by the same argument as above, we can show
that the vacuum oscillation solution is allowed,
as far as (\ref{eqn:max3}) is concerned\footnote{The critical
temperature for the vacuum oscillation solution is
$T\simeq 0.3{\rm MeV}$, so the effective degrees $g_\ast(T)$
of freedom becomes smaller than 10.75.  Nevertheless, it turns out
that $\max_{{}_T}~({\rm 3rd~term})\lsim{\cal O}(10^{-1})\times
(\Delta m_{21}^2/10^{-5}{\rm eV^2})^{1/2}$
is satisfied also for the
vacuum solution and this maximum value
is obviously much less than 1.}.
{}From (\ref{eqn:s13s14}) and (\ref{eqn:a}) it follows
\begin{eqnarray}
\vert U_{s3}\vert^2,\vert U_{s4}\vert^2\lsim 1.3\times 10^{-2}
\label{eqn:s31}
\end{eqnarray}
{}From (\ref{eqn:s23s24}) we have $0.33\lsim c_{24}^2\lsim 0.82$,
so that
\begin{eqnarray}
\vert U_{s4}\vert^2\simeq c_{24}^2s_{34}^2\gsim 0.33~s_{34}^2,
\label{eqn:s32}
\end{eqnarray}
where we have used the relations
\begin{eqnarray}
U_{s3}&=&c_{13}c_{34}s_{23}e^{-i\delta_1}
-c_{24}s_{13}s_{14}s_{34}e^{i(\delta_2-\delta_3)}
-c_{13}c_{23}s_{24}s_{34}\\
U_{s4}&=&c_{14}c_{24}s_{34}.
\end{eqnarray}
{}From (\ref{eqn:s31}) and (\ref{eqn:s32}) we conclude
\begin{eqnarray}
s_{34}^2\lsim 1.3\times 10^{-2}/0.33\simeq 3.8\times 10^{-2},
\end{eqnarray}
or
\begin{eqnarray}
\theta_{34}\lsim  11^\circ.
\label{eqn:th34}
\end{eqnarray}
{}From  (\ref{eqn:s13s14}) and (\ref{eqn:th34}), we have
\begin{eqnarray}
\vert U_{s3}\vert^2\simeq s_{23}^2\lsim 1.3\times 10^{-2}
\end{eqnarray}
or
\begin{eqnarray}
\theta_{23}\lsim  6^\circ.
\label{eqn:th23}
\end{eqnarray}

To summarize, for $N_\nu<4$ the mixing matrix looks like
\begin{eqnarray}
V_{KM}\sim\left(
\begin{array}{cccc}
c_\odot & s_\odot & \epsilon & \epsilon\\
\epsilon & \epsilon&c_{\rm atm} & s_{\rm atm} \\
\epsilon & \epsilon&-s_{\rm atm} & c_{\rm atm} \\
-s_\odot & c_\odot & \epsilon & \epsilon\\
\end{array}\right),
\label{eqn:mixingia}
\end{eqnarray}
where we have defined $\theta_\odot\equiv\theta_{12}$,
$\theta_{\rm atm}\equiv\theta_{24}$,
$\epsilon$ stands for a small number, and all the CP violating
phases have been dropped out because $|\theta_{13}|, |\theta_{14}|,
|\theta_{23}|, |\theta_{34}|$ $\ll 1$.
The basic reason that we have a strong constraint such as
(\ref{eqn:mixingia}) is because we have demanded that one of the
mass scales is given by $\Delta m_{\rm LSND}^2$, which
implies a severe bound on $\Gamma_{\nu_s}/H$ from big bang
nucleosynthesis.

It was pointed out in \cite{bbn1} that $\nu_e\leftrightarrow\nu_s$
oscillation is potentially dangerous because it could change the
density of $\nu_e$ at $T\lsim T_\nu$ and therefore could increase the
nucleon freeze-out temperature.  To avoid this effect, it is necessary
for the probability $P(\nu_e\rightarrow\nu_s)$ to satisfy
\begin{eqnarray}
{1 \over 2}\left(1+(1-P(\nu_e\rightarrow\nu_s))\right)
=\left({10.75 \over 10.75+ {7 \over 4}(N_\nu-3)}\right)^{1/2}
>0.93
\label{eqn:regionb1}
\end{eqnarray}
where we have substituted $G_F^2\rightarrow G_F^2
(1+(1-P(\nu_e\rightarrow\nu_s))/2$ in the equilibrium condition
\begin{eqnarray}
G_F^2T^5\simeq g_\ast(T)^{1/2}T^2/m_{\rm pl}
\label{eqn:equilibrium}
\end{eqnarray}
in case of $\nu_e\leftrightarrow\nu_s$ oscillation, while
we have used $N_\nu<3.9$ \cite{bbn4} and have put
$g_\ast(T)=10.75\rightarrow 10.75+(7/4)(N_\nu-3)
<10.75+(7/4)\times 0.9$ on the right hand side of (\ref{eqn:equilibrium}).
As we have seen above, the only case where
$\nu_e\leftrightarrow\nu_s$ oscillation occurs is
$T\sim 2$MeV (See (\ref{eqn:nusnue})), and using (\ref{eqn:regionb1})
and the condition $A\ll 1$ we have
\begin{eqnarray}
\sin^22(\theta_{12}^M+\theta_\odot)<0.26.
\label{eqn:regionb2}
\end{eqnarray}
The numerical value here differs slightly from the one in \cite{bbn1},
but in either case we arrive at the conclusion that the large-angle
MSW solution is excluded, since (\ref{eqn:th12m}) exceeds 0.26 for
some region of $T$ for $0.6\lsim\sin^22\theta_\odot\lsim 0.9$.  Note
that this argument does not apply to the vacuum solution, because the
collision length is much shorter than the oscillation length.  We note
in passing that the large-angle MSW solution for
$\nu_e\leftrightarrow\nu_s$ is excluded without referring to big bang
nucleosynthesis, by combining all the data of solar neutrino
experiments, the earth effect and Kamiokande day-night effect
\cite{solar3}\cite{solar4}.

It has been shown in the two flavor
analysis \cite{solar5} that the vacuum oscillation solution for the
channel $\nu_e\leftrightarrow\nu_s$ is excluded at 95 \% confidence
level if the standard solar model is taken for granted.  The mixing
matrix (\ref{eqn:mixingia}) tells us that our model in case of
$N_\nu<4$ is described approximately by two-flavor mixings
$\nu_e\leftrightarrow\nu_s$ and $\nu_\mu\leftrightarrow\nu_\tau$, so
we conclude that the vacuum oscillation solution is also excluded in
our scenario.  Hence we are left only with the small-angle MSW
solution for $N_\nu<4$.  This is consistent with a naive anticipation
that the small-angle MSW solution is the most preferable scenario,
because it fits the solar neutrino data best
\cite{solar2}--\cite{solar5}.

\section{Double $\beta$ Decay}

Assuming that neutrinos are of Majorana type, let us now consider the
implication to the neutrinoless double $\beta$ decay experiments.
To discuss neutrinoless double $\beta$ decays, we have to consider
the magnitudes of neutrino masses instead of mass squared differences.
{}From our assumption on the mass hierarchy
$\Delta m_{21}^2,\Delta m_{43}^2\ll\Delta m_{31}^2,\Delta m_{41}^2,
\Delta m_{32}^2,\Delta m_{42}^2$, we have
\begin{eqnarray}
m_1\simeq m_2,~
m_3\simeq m_4\simeq\sqrt{m_1^2+\Delta m_{31}^2}.
\label{eqn:mass}
\end{eqnarray}
{}From the form of the mixing matrix (\ref{eqn:mixingia})
the two pairs ($\nu_e, \nu_s$) and ($\nu_\mu, \nu_\tau$)
have basically no mixing with each other, and we have
\begin{eqnarray}
(m_{\nu_e},m_{\nu_s})\simeq m_1,&~&
(m_{\nu_\mu},m_{\nu_\tau})\simeq \sqrt{m_1^2+\Delta m_{31}^2}\nonumber\\
&{\ }&{\rm for~mass~pattern~(ia)}.
\label{eqn:massia}
\end{eqnarray}
In case of the mass pattern (\ref{eqn:patternib}), the mixing matrix
looks like
\begin{eqnarray}
V_{KM}\sim\left(
\begin{array}{cccc}
\epsilon & \epsilon&c_\odot & s_\odot  \\
c_{\rm atm} & s_{\rm atm}&\epsilon & \epsilon \\
-s_{\rm atm} & c_{\rm atm}&\epsilon & \epsilon \\
\epsilon & \epsilon&-s_\odot & c_\odot  \\
\end{array}\right),
\label{eqn:mixingib}
\end{eqnarray}
so that we have
\begin{eqnarray}
(m_{\nu_e},m_{\nu_s})\simeq \sqrt{m_1^2+\Delta m_{31}^2},&~&
(m_{\nu_\mu},m_{\nu_\tau})\simeq m_1\nonumber\\
&{\ }&{\rm for~mass~pattern~(ib)}.
\label{eqn:massib}
\end{eqnarray}

It is well known \cite{bhp}--\cite{fy} that
the CP violating phases $\alpha,\beta,\gamma$ in (\ref{eqn:u})
cannot be absorbed by redefinition of the neutrino fields.  So the
effective electron neutrino mass, which is the
e-e component of the mass matrix, is given by
\begin{eqnarray}
\langle m_{\nu_e}\rangle&\equiv&\left\vert\left[
V_{KM}~D~{\rm diag}\left(m_j\right)~D~V_{KM}^T\right]_{ee}
\right\vert\nonumber\\
&\simeq&\left\{ \begin{array}{lr}
m_1\left\vert e^{i\alpha}c_\odot^2+e^{-i\alpha}s_\odot^2
\right\vert\simeq m_1\nonumber\\
\qquad\hfill{\rm for~mass~pattern~(ia)}~\nonumber\\
\nonumber\\
\sqrt{m_1^2+\Delta m_{31}^2}\left\vert e^{i\alpha}
c_\odot^2+e^{-i\alpha}s_\odot^2
\right\vert\simeq \sqrt{m_1^2+\Delta m_{31}^2}\nonumber\\
\qquad\hfill{\rm for~mass~pattern~(ib)},\\
\end{array} \right.\\
\label{eqn:mnue1}
\end{eqnarray}
where $D\equiv e^{-i\sqrt{6}\gamma\lambda_{15}}e^{-i\beta\lambda_8}
e^{-i\alpha\gamma\lambda_3}$ is a diagonal matrix and we have used
(\ref{eqn:massia}), (\ref{eqn:massib}) and the condition
$\sin^22\theta_\odot\simeq 6\times 10^{-3}$ for the small-angle MSW
solution.  The present upper bound on $\langle m_{\nu_e}\rangle$ from
neutrinoless double $\beta$ decay experiments is
\cite{moe}\cite{zuber}
\begin{eqnarray}
\langle m_{\nu_e}\rangle < 0.68 {\rm eV}.
\label{eqn:mnue2}
\end{eqnarray}
{}From (\ref{eqn:mnue1}), (\ref{eqn:mnue2}), (\ref{eqn:massia})
and (\ref{eqn:massib}) we obtain
\begin{eqnarray}
m_1&\lsim& 0.68 {\rm eV}
\qquad\qquad{\rm for~mass~pattern~(ia)},
\label{eqn:mnueia}\\
\nonumber\\
m_1&\lsim& \sqrt{\left(0.68 {\rm eV}\right)^2-\Delta m_{31}^2}\nonumber\\
&\lsim&0.44 {\rm eV}
\qquad\qquad{\rm for~mass~pattern~(ib)},
\label{eqn:mnueib}
\end{eqnarray}
where in case of (\ref{eqn:mnueib}) we have used the lower bound in
(\ref{eqn:delmlsnd}).  Note that
$\Delta m_{\rm LSND}^2\simeq
\Delta m_{31}^2 < 0.46 {\rm eV}^2$
has to be satisfied in (\ref{eqn:mnueib}) for $m_1$ to be real.

\section{Hot Dark Matter}

It has been argued that a model with cold+hot dark matter with
5eV$\lsim\sum_\alpha m_{\nu_\alpha}$ $\lsim$ 7eV seems
to be in agreement with observations, such as the anisotropy of the
cosmic background radiations, correlation of galactic clusters, etc.
\cite{phkc}\cite{dm}, and efforts have been made to introduce sterile
neutrinos to account for hot dark matter as well as other anomalies
\cite{pv}--\cite{p}.
Here we examine
the possibility that neutrinos could be
hot dark matter while satisfying all the constraints
that we have obtained in the previous sections.

Since we assume that sterile neutrinos have never been in thermal
equilibrium, only three components of the mass matrix
(\ref{eqn:massia}) or (\ref{eqn:massib})
contribute to the mass density.  Using
(\ref{eqn:mnueia}) and (\ref{eqn:mnueib}), therefore, we have the following
total mass bound:
\begin{eqnarray}
m_{\nu_e}+m_{\nu_\mu}+m_{\nu_\tau}
\simeq \left\{ \begin{array}{lr}
m_1+2\sqrt{m_1^2+\Delta m_{31}^2}\lsim 4.0{\rm eV}\\
\qquad\hfill{\rm for~mass~pattern~(ia)}~\nonumber\\
\nonumber\\
2m_1+\sqrt{m_1^2+\Delta m_{31}^2}\lsim 1.6{\rm eV}\\
\qquad\hfill{\rm for~mass~pattern~(ib)}.\\
\end{array} \right.\\
\label{eqn:totalmassia}
\end{eqnarray}
{}From (\ref{eqn:totalmassia})
we conclude that in either case neutrinos are not heavy enough
to account for all the hot dark matter components.

\section{$N_\nu \ge 4 $}

The possibility of $N_\nu \ge 4 $ has been proposed recently
\cite{ks}, so we will also discuss this case for the sake of
completeness.  If $N_\nu \ge 4 $, then sterile neutrinos should have
been in thermal equilibrium, and we cannot put a strong constraint
like the one for $N_\nu < 4 $.  The only condition which has to be
satisfied for sterile neutrinos to be in thermal equilibrium
is $A>1.3\times10^{-2}$.  We still have
constraints from the reactor experiments which imply that
$\theta_{13},\theta_{14}$ are small, but we can no longer say anything
about the magnitude of $\theta_{23},\theta_{34}$.  The large-angle MSW
solution is allowed in this case, as long as $\nu_s$ is in thermal
equilibrium.  As we noted earlier, however, the large-angle MSW and
vacuum oscillation solutions are excluded for
$\nu_e\leftrightarrow\nu_s$ in the two flavor analysis
\cite{solar3}\cite{solar4}\cite{solar5}, so if $U$ is very close to a
direct sum of two mixing matrices with $\nu_e\leftrightarrow\nu_s$ and
$\nu_\mu\leftrightarrow\nu_\tau$ channels then it contradicts with solar
neutrino observations.  The analysis of solar neutrino problem with
four species of neutrinos would be extremely complicated\footnote{The
explicit calculation for solar neutrino problem with three flavors has
been performed recently in \cite{flm}, assuming mass hierarchy.}, and
it is yet to be seen under what conditions of mixing angles the
large-angle MSW and vacuum oscillation solutions are allowed.

Here, for simplicity, let us consider an extreme case in which the
oscillations take place mainly in
the channels $\nu_e\leftrightarrow\nu_\tau$ and
$\nu_\mu\leftrightarrow\nu_s$.  This solution is consistent with all
the solar neutrino observations
\cite{solar3}\cite{solar4}\cite{solar5}, and $\nu_s$ is in thermal
equilibrium because the large mixing angle of
$\nu_\mu\leftrightarrow\nu_s$ suggested by the atmospheric neutrino
anomaly \cite{kamioka2} gives $\Gamma_{\nu_s}/H$ which is larger than
one \cite{cf2}.  The masses of neutrinos in this case are given by
\begin{eqnarray}
&{\ }&(m_{\nu_e},m_{\nu_\tau})\simeq m_1,~
(m_{\nu_\mu},m_{\nu_s})\simeq \sqrt{m_1^2+\Delta m_{31}^2}\nonumber\\
&{\ }&\qquad{\rm for~mass~pattern~(ia)},
\label{eqn:massia4}
\end{eqnarray}
\begin{eqnarray}
&{\ }&(m_{\nu_e},m_{\nu_\tau})\simeq \sqrt{m_1^2+\Delta m_{31}^2},~
(m_{\nu_\mu},m_{\nu_s})\simeq m_1\nonumber\\
&{\ }&\qquad{\rm for~mass~pattern~(ib)}.
\label{eqn:massib4}
\end{eqnarray}

If the solar neutrino problem is solved by the small-angle MSW
solution, then from (\ref{eqn:mnueia}) and (\ref{eqn:mnueib})
we have
\begin{eqnarray}
m_{\nu_e}+m_{\nu_\mu}+m_{\nu_\tau}+m_{\nu_s}
&\simeq& 2m_1+2\sqrt{m_1^2+\Delta m_{31}^2}\nonumber\\
&\lsim&\left\{ \begin{array}{lr}
4.7{\rm eV}\qquad
{\rm for~mass~pattern~(ia)}~\nonumber\\
2.2{\rm eV}\qquad
{\rm for~mass~pattern~(ib)}.\\
\end{array} \right.\\
\label{eqn:totalmassib4s}
\end{eqnarray}
So in this case neutrinos cannot account for all the hot dark matter
components.

If we take the large-angle MSW solution, on the other hand,
instead of (\ref{eqn:mnue1}) we have
\begin{eqnarray}
\langle m_{\nu_e}\rangle
\simeq\left\{ \begin{array}{ll}
m_1\sqrt{1-\sin^2\alpha\sin^22\theta_\odot}\,
\gsim\, 0.32m_1\\
\qquad\hfill{\rm for~mass~pattern~(ia)}~\\
\nonumber\\
\sqrt{m_1^2+\Delta m_{31}^2}
\sqrt{1-\sin^2\alpha\sin^22\theta_\odot}
\gsim\, 0.32\sqrt{m_1^2+\Delta m_{31}^2}\\
\qquad\hfill{\rm for~mass~pattern~(ib)},\\
\end{array} \right.\\
\label{eqn:mnue4l}
\end{eqnarray}
where we have used the condition $0.6\lsim\sin^22\theta_\odot\lsim 0.9$
for the large-angle MSW solution.  From (\ref{eqn:mnue2}) and
(\ref{eqn:mnue4l}) we have
\begin{eqnarray}
\left\{ \begin{array}{rlll}
m_1&\lsim& 2.2{\rm eV}&{\ }\nonumber\\
2m_1+2\sqrt{m_1^2+\Delta m_{31}^2}
&\lsim&9.6{\rm eV}&\qquad{\rm for~mass~pattern~(ia)},\nonumber\\
\end{array} \right.\\
\\
\left\{ \begin{array}{rlll}
m_1&\lsim& 2.1{\rm eV}&{\ }\nonumber\\
2m_1+2\sqrt{m_1^2+\Delta m_{31}^2}
&\lsim&8.5{\rm eV}&\qquad{\rm for~mass~pattern~(ib)},\nonumber\\
\end{array} \right.\\
\end{eqnarray}
so that neutrinos could explain all the hot dark matter.
Conversely, if all the hot dark matter components are neutrinos
in this case, then it follows
\begin{eqnarray}
2m_1+2\sqrt{m_1^2+\Delta m_{31}^2}
&\gsim&5{\rm eV},\\
m_1&\gsim& 0.8{\rm eV},\\
\langle m_{\nu_e}\rangle&\gsim&
\left\{ \begin{array}{rlll}
0.2{\rm eV}&\qquad{\rm for~mass~pattern~(ia)}\nonumber\\
0.4{\rm eV}&\qquad{\rm for~mass~pattern~(ib)},\nonumber\\
\end{array} \right.\\
\label{eqn:totalmass4l}
\end{eqnarray}
so we will be able to observe neutrinoless double $\beta$ decays
in the future experiments.

If the solar neutrino problem is solved by the vacuum oscillation
solution, we have
\begin{eqnarray}
\langle m_{\nu_e}\rangle
\simeq \left\{ \begin{array}{lr}
m_1\sqrt{1-\sin^2\alpha\sin^22\theta_\odot}
\ge 0,\\
\qquad\hfill{\rm for~mass~pattern~(ia)}~\nonumber\\
\nonumber\\
\sqrt{m_1^2+\Delta m_{31}^2}
\sqrt{1-\sin^2\alpha\sin^22\theta_\odot}\,
\ge\, 0\\
\qquad\hfill{\rm for~mass~pattern~(ib)},\\
\end{array} \right.\\
\label{eqn:mnue4v}
\end{eqnarray}
where we have used the condition $0.6\lsim\sin^22\theta_\odot\le 1$
for the vacuum oscillation solution.  If $\alpha$ and $2\theta_\odot$
are both very close to $\pi/2$, then $m_1$ can be arbitrarily large
without contradicting with the bound (\ref{eqn:mnue2}) from the
neutrinoless double $\beta$ decay experiments.  In this case,
therefore, neutrinos could account for all the hot dark matter
components, but there is no guarantee that we will be able to observe
neutrinoless double $\beta$ decays in future experiments.

\section{Conclusions}

In this paper we have performed a detailed analysis of the constraints
on a model with three active and one sterile neutrinos, using the data
of reactor and accelerator experiments, the solar and the atmospheric
neutrino observations, and big bang nucleosynthesis.  The mass pattern
where three masses are degenerated is found to be inconsistent with
reactor and accelerator experiments.  If $N_\nu<4$, then all the
mixing angles are severely constrained and the mixing matrix is
effectively split into 2$\times$2 matrices with channels
$\nu_e\leftrightarrow\nu_s$ and $\nu_\mu\leftrightarrow\nu_\tau$.  In
this case the large-angle MSW and vacuum oscillation solutions are
excluded.  Because of the constraints from neutrinoless double $\beta$
decay experiments, neutrinos cannot explain all the hot dark matter
components.  For $N_\nu\ge 4$, we get fewer conditions on the mixing
angles, and all the solutions to the solar neutrino problem are
allowed.  If we take either the large-angle MSW solution or the vacuum
oscillation one, then these neutrinos can account for all the hot dark
matter components.  In case of the large-angle MSW solution for
$N_\nu\ge 4$, we will be able to observe neutrinoless double $\beta$
decays in near future.  It is hoped that combined results of
super-Kamiokande and SNO experiments can tell us about the existence
of sterile neutrinos, the type of the solution to the solar neutrino
problem, and the mixing of sterile neutrinos \cite{bg}.

\end{document}